\documentclass[manuscript]{acmart}
\AtBeginDocument{%
  \providecommand\BibTeX{{%
    \normalfont B\kern-0.5em{\scshape i\kern-0.25em b}\kern-0.8em\TeX}}}

\setcopyright{acmcopyright}
\copyrightyear{2022}
\acmYear{2022}
\acmDOI{XXXXXXX.XXXXXXX}

\acmConference[Conference acronym 'XX]{Make sure to enter the correct
  conference title from your rights confirmation emai}{June 03--05,
  2022}{Woodstock, NY}
\acmPrice{15.00}
\acmISBN{978-1-4503-XXXX-X/18/06}

\usepackage{multirow}
\usepackage{array}
\usepackage{booktabs}
\usepackage{float} 
\usepackage{subfigure}
\begin{document}

\title{Context-aware explainable recommendations over knowledge graphs}

\author{Jinfeng ZHONG}
\affiliation{%
  \institution{Paris-Dauphine University, PSL Research University, CNRS UMR 7243, LAMSADE}
  \city{Paris}
  \country{France}
  \postcode{75016}}
\email{jinfeng.zhong@dauphine.eu}

\author{Elsa NEGRE}
\affiliation{%
  \institution{Paris-Dauphine University, PSL Research University, CNRS UMR 7243, LAMSADE}
  \city{Paris}
  \country{France}
  \postcode{75016}}
\email{elsa.negre@lamsade.dauphine.fr}
\renewcommand{\shortauthors}{Zhong and Negre}

\begin{abstract}
 Knowledge graphs contain rich semantic relationships related to items and incorporating such semantic relationships into recommender systems helps to explore the latent connections of items, thus improving the accuracy of prediction and enhancing the explainability of recommendations. However, such explainability is not adapted to users' contexts, which can significantly influence their preferences. In this work, we propose CA-KGCN (\textbf{C}ontext-\textbf{A}ware \textbf{K}nowledge \textbf{G}raph \textbf{C}onvolutional \textbf{N}etwork), an end-to-end framework that can model users' preferences adapted to their contexts and can incorporate the rich semantic relationships in the knowledge graph related to items. This framework captures users' attention to different factors: contexts and features of items. More specifically, the framework can model users' preferences adapted to their contexts and provide explanations adapted to the given context. Experiments on three real-world datasets show the effectiveness of our framework: modeling users' preferences adapted to their contexts and explaining the recommendations generated.
\end{abstract}



\maketitle

\section{Introduction}\label{sec:intro}

Recommender systems (RSs hereinafter) have become important tools to alleviate information overload and can help users find the items or services that are really of their interest more quickly \cite{zhang2020explainable}. Data records in traditional RSs are in the form of matrix or tensors \cite{karatzoglou2010multiverse}. Therefore, traditional recommendation approaches can be resumed as matrix completion problem. In recent years, miscellaneous information resources such as users' contexts, knowledge graphs (KGs hereinafter) have become available for modeling users' preferences. However, modeling such data in the form of tensors suffers from high complexity. Researchers have turned to graph-based approaches to model such complex data. In context-aware recommender systems (CARSs hereinafter), data records in CARSs are usually represented in the form of $<user, item, contexts, rating>$ \cite{adomavicius2011context}, which can also be represented by a 3-partite graph (users, items and contexts are the three kinds of nodes in the graph) \cite{duran2021graph}. KGs are defined as heterogeneous graphs \emph{intended to accumulate and convey
knowledge of the real world} \cite{hogan2021knowledge}. Leveraging KGs helps to enhance the quality \cite{guo2020survey} and explainability \cite{wang2018ripplenet, wang2019knowledge,wang2019knowledge1} of recommendations. In these cases, recommendation methods can be abstracted as a graph exploration problem, which is also the case in many other recommendation scenarios \cite{kyriakidi2020recommendations}.

In another line of research, explainable recommendations have become an increasingly popular topic in industry and academia \cite{zhang2020explainable}, which requires that RSs not only recommend personalized items but also explain why such recommendations are generated. Explanation methods in RSs can be classified by the information sources \cite{zhang2020explainable}: relevant users or items, item features, social relationships, etc. Thus, these explanations are not adapted to users' contexts. Since users' preferences vary across different contexts, explanations should also be context-aware \cite{zhang2020explainable,zhong2022towards}.

As we presented above, user-item-contexts interactions can naturally be represented in a 3-partite graph; KGs are heterogeneous graphs. The key challenge is how to model such graph-based data. The most prominent technique for learning graph representation is Graph Convolutional Network (GCN hereinafter) \cite{kipf2016semi}. GCN has been demonstrated to be effective in generating context-aware recommendations \cite{duran2021graph, wu2020graph}. In \cite{duran2021graph}, contexts are modeled as nodes while in \cite{wu2020graph}, contexts are represented as edges connecting users and items. Both works apply message propagation to refine the representation of users, items, and contexts. However, both works do not leverage KGs and do not involve the explainability of recommendations. GCN has also shown its effectiveness in learning the embeddings of entities and relations in KGs \cite{dadoun2019location, wang2019multi, wang2019knowledge, wang2019knowledge1}. In these works, items and their features correspond to nodes; relations correspond to edges. However, users' contexts are not considered, which means that the recommendations are not adapted to their contexts.

\begin{figure}[t]
\centering
\includegraphics[scale=0.29]{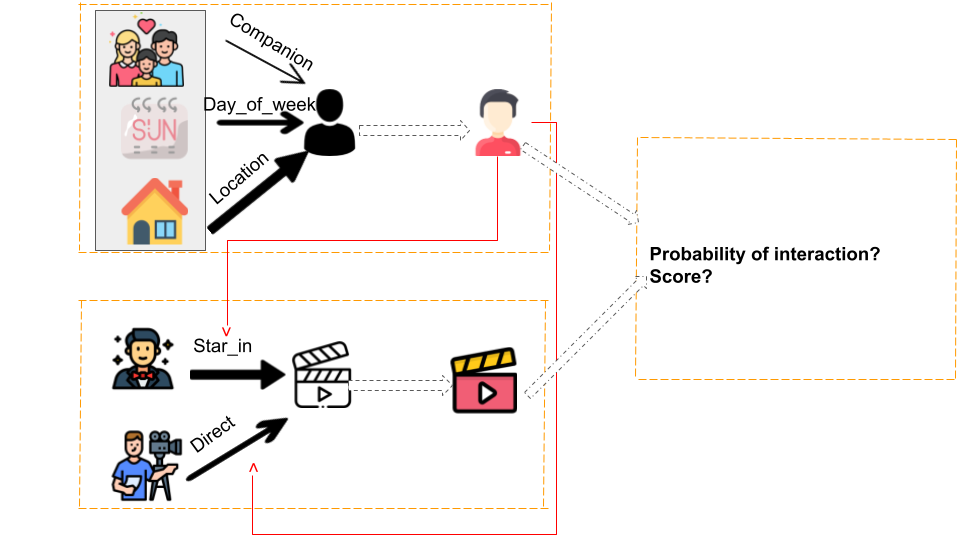}
\caption{A toy example that describes how contextual situations reshape users' preferences, how KG related to items helps to capture users' preferences. The solid arrows represent relations in knowledge and contextual factors. The thicker they are, the more important are the relations and contextual factors for this specific user. }
\label{fig:toy_ex}
\vspace{-0,4cm}
\end{figure}

In this work, we wish to fill this gap by leveraging users' contexts and KGs at the same time to generate and explain recommendations. The basic idea of this framework is: when choosing items, users pay different levels of attention to different factors. Figure~\ref{fig:toy_ex} shows this idea by a toy example in movie recommendation. This toy example monitors the decision-making process in movie recommendation. When choosing a movie, some users are more likely to be influenced by \textbf{\emph{companion}} (e.g. with lovers or with children) while others may more likely to be influenced by \textbf{\emph{location}} (e.g. home or public place). At the same time, when making a decision, some users may have special preferences for certain actors while others may prefer movies by certain directors. In the first case, these users pay higher attention to the relation \textbf{\emph{stares\_in}} while in the second case users accord more importance to the relation \textbf{\emph{directs}}.

To model such differences, we propose Context-Aware Knowledge Graph Convolutional Network (CA-KGCN hereinafter). In particular, this framework is composed of three components: \emph{user-context embedding layer}, \emph{knowledge graph embedding layer} and \emph{output layer}. Specifically, the \emph{user-context embedding layer} computes the representation of users under different contexts. Based on users' representation under different contexts, the \emph{knowledge graph embedding layer} first computes the importance of relations in the knowledge graph related to items, which refines the representation of items under the same context. Lastly, the \emph{output layer} computes the interaction probabilities of user-item pairs or the potential score users may give to items. To summarize, our framework has the following traits: (1) CA-KGCN can leverage users' contexts and the knowledge graph related to items, enabling it to accurately model users' preferences according to their contexts and generate context-aware recommendations; (2) CA-KGCN can model the importance that users accord to their contexts and item features, which helps to generate personalized and context-aware explanations. 

We summarize the contributions of our work as follows: (1) We highlight the importance of generating context-aware explanations in RSs; (2) We propose CA-KGCN (Context-Aware Knowledge Graph Convolutional Network), a framework that can combine users' contexts and KGs to generate personalized recommendations adapted to users' contexts; (3) We conduct experiments on three real-world datasets to verify the effectiveness of our model; (4) Through a case study we show that our framework can identify the most influential relations in KGs and contexts, which helps to generate context-aware explanations.

\section{Related work} \label{sec:related_work}

In this section, we will first lay out the preliminary definitions used in this paper. Then we review related work involved in this paper: explanations in RSs, context-aware recommender systems and GCN.

\subsection{Preliminary} \label{sec:Preliminary}
In this section, we introduce the relevant concepts and definitions used in this paper. The important definitions used in this paper include the following:

 \textbf{Knowledge graphs:} Knowledge graphs are directed heterogeneous graphs \emph{intended to accumulate and convey knowledge of the real world} \cite{hogan2021knowledge}. More specifically, items and attributes of items correspond to nodes; relations correspond to edges. Formally, a knowledge graph is defined as following: $\mathcal{G = (\mathcal{E},R)}$, where $\mathcal{E} = \{e_1,e_2,\dots\}$ is a set of entities in $\mathcal{G}$ and $\mathcal{R} = \{r_1,r_2,\dots\}$ is a set of relations in $\mathcal{G}$. $\mathcal{G}$ is composed of \emph{entity-relation} triplets $(e_1, r, e_2)$, where $e_1 \in \mathcal{E}$, $r \in \mathcal{R}$, $e_2 \in \mathcal{E}\backslash e_1$. Here is a concrete example in a phone application recommendation scenario, a triplet can be $(Facebook, category, social)$, indicating the fact that \emph{Facebook} is a social mobile app. 
 
 \textbf{Context:} In this paper we adopt the definition proposed by Dey et al. \cite{abowd1999towards}: ``Context is any information that can be used to characterize the situation of an entity.'' A contextual situation $cs$ is composed of several contextual conditions. As a concrete example, in Figure~\ref{fig:toy_ex}, the situation of the user is characterized by contextual information: ``Companion'', ``Day of the week'' and ``Location'', which we term as contextual factors. Each contextual factor can have several possible values (these values are noted as contextual conditions). For example, the possible values of the contextual factor ``Companion'' include: ``With family'', ``With friends'', ``With lover'', ``With colleague''. A contextual situation is composed of several contextual conditions. Therefore, the situation of the user in Figure\ref{fig:toy_ex} is $(sunday, at\_home, with\_family)$. In the rest of the paper, we equate \emph{context} and \emph{contextual situation}.
 
 \subsection{Explanations in recommender systems}

Recent years have seen the proliferation of research in explaining recommendations \cite{zhang2020explainable}. This is because explaining recommendations can help users find the items that are really of their interest (effectiveness) more quickly (efficiency), convince users to try or to consume (persuasiveness), make the systems more transparent (transparency), increase users' loyalty in the system (trust) and increase the ease of use or enjoyment (satisfaction) \cite{tintarev2015explaining}. Besides, explaining recommendations help developers to debug and improve the performances of RSs \cite{lipton2018mythos}. Therefore, it is necessary to provide explanations to justify recommendations.

Miscellaneous information sources have been used to explain recommendations, which vary across different RSs \cite{zhang2020explainable}. Relevant users or items are usually utilized in collaborative filtering models to explain recommendations \cite{sarwar2001item, cleger2012explaining}; features of items are typically used to explain recommendations in content-based RSs \cite{vig2009tagsplanations}; social relationships are leveraged to explain recommendations in RSs that explore users' social relationships \cite{quijano2017make, park2017uniwalk}; opinion-based explanations typically leverage users' comments on items or items' features \cite{mcauley2013hidden}. However, these explanations are static, which means that the explanations returned by these methods remain the same under different contextual situations for the same users. Usually, the corresponding RS ignores the influences of contexts. As a matter of fact, contexts can influence users' preferences \cite{adomavicius2011context}. Since users' preferences vary across contexts, generating explanations adapted to users' contextual situations would be an appropriate way to explain recommendations \cite{zhang2020explainable, zhong2022towards}. We hence explore context-aware explanations in this paper.

\subsection{Context-aware recommender systems}
CARSs leverage users' contexts to model users' preferences with finer granularity, therefore CARSs can provide more personalized recommendations adapted to their contextual situations (e.g. companion, time and location, etc.). From a chronological point of view, tensor factorization \cite{karatzoglou2010multiverse} methods were first applied to model the user-item-contexts data, which is an extension of matrix factorization (MF hereinafter) \cite{koren2009matrix}. Factorization machine (FM hereinafter) \cite{rendle2010factorization} further generalizes the idea of MF to model the pair-wise relationships of features. FM has been used to generate context-aware recommendations because it can capture the second-order interactions between users, items and contexts. There are also attempts that combine neural networks to further enhance the performances of FM: Neural factorization machines \cite{he2017neural}, DeepFM \cite{ guo2017deepfm}, xdeepfm \cite{lian2018xdeepfm}.

More recently, GCNs are applied to capture the high order interactions among users, items and contexts because the data recods in CARSs can be represented as a 3-partite graph. In \cite{duran2021graph}, data records are represented by user-item-contexts interaction tensors equivalent to a 3-partite graph, then embedding propagation aggregates the embedding of neighboring nodes iteratively to refine the representation of source nodes. In \cite{wu2020graph}, users and items are represented as nodes, contexts are represented as edges. Users and items are first cast into a embedding space by an encoder that incorporates theirs features then a GCN layer incorporates information from contexts into users and items, lastly a decoder layer outputs the prediction scores. Similar to \cite{duran2021graph}, FM is utilized to output the prediction scores. Knowledge graph is not leveraged in \cite{duran2021graph,wu2020graph} and explainability of recommendations is not involved in \cite{duran2021graph}, neither in \cite{wu2020graph}. In this paper we propose CA-KGCN that can generate context-aware recommendations and corresponding context-aware explanations in RSs.

\subsection{Graph convolutional networks }
    To deal with the emerging graph-based data, graph neural networks have emerged in many areas and GCN is among the most prominent graph neural networks \cite{zhang2019graph}. GCN has been developed to extract the localized features on such graph-based data. The basic idea of GCN is that the representation of a node in a graph depends on its neighboring nodes and itself \cite{gao2021graph}, which means that GCN repeatedly aggregates the information from neighboring nodes with information of itself. GCN has been widely utilized in RSs for its power in learning graph representation, capturing high-order connectivity and, in modeling supervision signal and in leveraging semi-supervised signals \cite{jin2020multi,gao2021graph, zhang2019graph}.

GC-MC (Graph Convolutional Matrix Completion) \cite{berg2017graph} and LightGCN \cite{he2020lightgcn} model user-item interactions in a bipartite graph, which can capture the high-order interactions among users and items. Fi-GNN \cite{li2019fi} represents items in the form of graph to model the features interactions and is used in CTR (Click Through Rate) prediction tasks. Social relations can naturally be represented in graphs, nodes representing users and edges representing relationships. Therefore, GCN has also been employed in social recommendations \cite{fan2019graph}. Moreover, recent years have seen the proliferation of GCN in context-aware recommendations \cite{duran2021graph, wu2020graph} and in RSs that incorporates KGs \cite{wang2019knowledge, wang2019kgat}. Results in \cite{wang2019knowledge, wang2019knowledge1, wang2019kgat} show that leveraging knowledge graph in RSs can enhance the explainability of recommendations. This is because KGs contain rich semantics in entities and relations that compose the KGs. Considering GCN's ability to aggregate information from neighboring nodes in a graph, we believe that GCN can capture the interactions among users, items, contexts, entities and relations in knowledge graph. Therefore, more refined representation of users, items, contexts, entities and relations in knowledge graph can be achieved. The above works we mention either only leverage users' context or only knowledge graph. In our work, we aim to combine context and knowledge graph with the help of GCN to generate context-aware recommendations and context-aware explanations.
 
 \section{Our approach}
    In this section, we present the CA-KGCN (Context-Aware Knowledge Graph Convolutional Network) framework. We will first formulate the recommendation problem in CA-KGCN. Then we introduce the three components of CA-KGCN in detail. Note that in the rest of this paper, $u, i,cf, cd  \in \mathbb{R}^d$ indicate vectors representing user, item, contextual factor, contextual condition respectively; $ r, e \in \mathbb{R}^d$ indicate relation and entity in $\mathcal{G}$ respectively; $\pi_u^{cf}$ indicates the importance of contextual factor $cf$ to user $u$; $\pi_u^r$ indicates the importance of relation $r$ to user $u$.  

\subsection{Problem formulation} \label{sec:problem_formulation}
Suppose that there are $m$ users, $n$ items, each observation in a dataset is in the form of $\{u,i,cs, \mathcal{M}_{u,i,cs}\}$, where $\mathcal{M}_{u,i,cs}$ is the interaction matrix of users, items, contexts. Given the interaction matrix $\mathcal{M}_{u,i,cs}$ as well as the knowledge graph $\mathcal{G}$ related to items, the problem to solve becomes predicting the probability that user $u$ will interact with item $i$ or predicting the score the user $u$ will potentially give to item $i$ under a target contextual situation $cs$. More formally, the goal is to learn a prediction function $f: u, i, \mathcal{G}, cs \rightarrow \mathbb{R}$ that can predict such probability (score). 

\subsection{Methodology}\label{sec:methodology}
\begin{figure*}[t]
\centering
\includegraphics[scale=0.269]{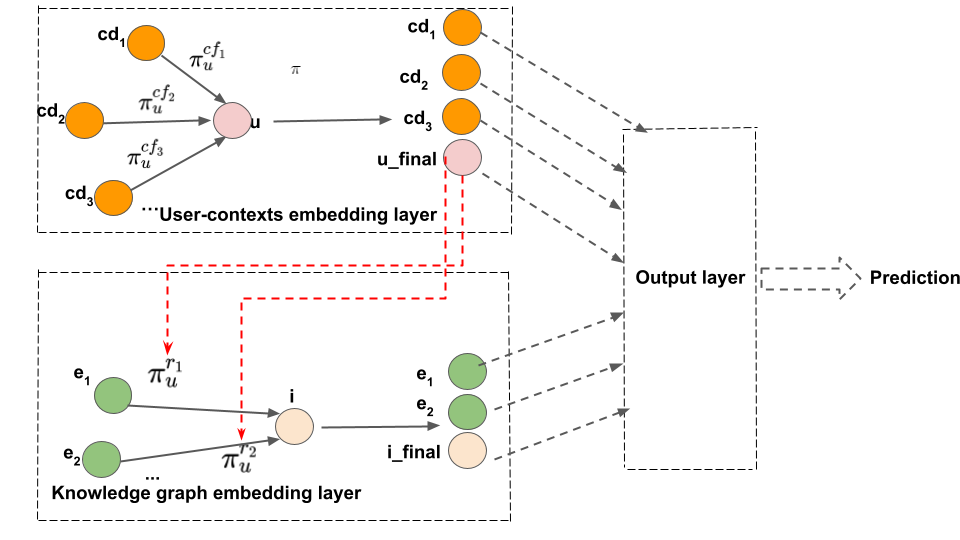}
\caption{Illustration of context-aware knowledge graph convolutional network. }
\label{fig:framework}
\vspace{-0,4cm}
\end{figure*}

 Figure~\ref{fig:framework} illustrates the framework of our model that contains three components: a \emph{user-context embedding layer} followed by a \emph{knowledge graph embedding layer} and an \emph{output layer}. The \emph{user-context embedding layer} captures the influence of contexts on users to get a representation of users adapted to the target contextual situation. The \emph{knowledge graph embedding layer} refines the representation of items under the target contextual situation and based on the representation of users. The \emph{output layer} computes the predicted score or the interaction probabilities. In what follows, we will present the three components in order. 

\subsubsection{User-context embedding layer}\label{sec:user-context}
As presented in Figure~\ref{fig:framework}, the final representation of a user depends on the contextual situation of the user. In this step, we aim to compute the representation of users adapted to their contextual situation. To this end, we first calculate the importance of contextual factor $cf$ to a user $u$: 

\begin{equation}
    \pi_u^{cf} = g(u,cf)
\end{equation}

where $g$ is a inner production like in \cite{wang2019knowledge,wang2019knowledge1} where the authors used this function to compute the importance of relations in knowledge graphs to items.

Having calculated the importance $\pi_u^{cf}$ of all contextual factors to this user, we normalize the importance with Equation~\ref{equa:normalize} like in \cite{velivckovic2017graph}:

\begin{equation}\label{equa:normalize}
   \beta_u^{cf} = \frac{exp(LeakyReLU(\pi_u^{cf}))}{\sum_{cf \in C}exp(LeakyReLU(\pi_u^{cf}))}
\end{equation}

In sequence, we compute the representation of the contextual situation $cs$ by summing up all the vectors representing contextual conditions multiplied by the normalized importance. 
     \begin{equation}
         cs = \sum\limits_{cd \in cs} \beta _u^{cf}cd
     \end{equation}
     
The next step is aggregating the representation of contextual situation $cs$ with the representation of user $u$ to get a specific representation of user $u$ under the contextual situation $cs$. To this end, we user two types of aggregators like in \cite{wang2019knowledge1}: 

\begin{itemize}
    \item \textbf{Sum aggregator} that simply sums $u$ and $cs$.
    
    \begin{equation}
        u\_final\_sum = \sigma(w_1(u + cs) + b1)
    \end{equation}
     where $w_1$ and $b_1$ are transformation matrix and bias respectively, $\sigma$ is a non-linear activation function such as \emph{ReLU} \cite{paszke2019pytorch}.
     \item \textbf{Concat aggregator} that concatenates $u$ and $cs$. 
     
     \begin{equation}\label{equa:concat_c}
         u\_final\_cat = \sigma(w_2 * concat (cs,u)) + b_2
     \end{equation}
     where $w_2$ and $b_2$ are transformation matrix and bias respectively. This operation is to ensure that the representation of $u$ after concatenation operation is still of dimension $d$.
\end{itemize}

By aggregating information from a contextual situation $cs$ and a user $u$, each user $u$  gets a specific representation $u\_final$ under a target contextual situation $cs$. We will compare the effects of the two aggregators in Section~\ref{sec:ablation}.

\subsubsection{Knowledge graph embedding layer}\label{sec:knowledge-embedding}

This layer computes the refined representation of items by leveraging the semantics in the knowledge graph related to items. Considering the same user $u$ under the contextual situation $cs$, for a candidate item $i$, like in the \emph{user-context embedding layer}, we first calculate the importance of each relation $r$ in $\mathcal{G}$ to user $u$ by applying Equation~\ref{equa: impor_r}. Recall that the relation $r$ here corresponds to the relations in the knowledge graph related to item. Then, by applying the same normalization function presented by Equation~\ref{equa:normalize}, each $\pi_u^{r}$ is transformed into $\beta_u^{r}$.

\begin{equation} \label{equa: impor_r}
    \pi_u^{r} = g(u,r)
\end{equation}
where $u$ corresponds to the user representation obtained in the \emph{user-context embedding layer}, see Section~\ref{sec:user-context} for more detail.

Nodes in a knowledge graph can influence the states of neighboring nodes through message passing \cite{berg2017graph}. Therefore, to characterize the state of a node representing an item $i$, it is necessary that the states of its neighbors are aggregated. To get a complete representation of $i$'s neighbors, the vectors  representing its neighbors $e \in \mathcal{E}_i$ are summed and we term it as $\mathcal{E}_i$.

\begin{equation}
    \mathcal{E}_i = \sum\limits_{e \in \mathcal{E}_i} \beta _u^{r}e
\end{equation}

Like in the \emph{user-context embedding layer}, we use two aggregators to aggregate the representations of $i$'s neighboring entities with the representation of item $i$ in two ways:

\begin{itemize}
    \item \textbf{Sum aggregator} that simply sums up the representation of $i$'s neighboring entities and $i$. 
     \begin{equation}
         i\_final\_sum = \sigma(w_3(\mathcal{E}_i + i) + b_3)
     \end{equation}
     where $w_3$ and $b_3$ are transformation matrix and bias respectively.
     \item \textbf{Concat aggregator:} that concatenates $\mathcal{E}_i$ and i. 
     
     \begin{equation}\label{equa:concat_k}
         i\_final\_cat =\sigma( w_4 * concat(\mathcal{E}_i, i) + b_4)
     \end{equation}
     where $w_4$ and $b_4$ are transformation matrix and bias respectively. They ensure that the representation of $i$ after concatenation operation is still of dimension $d$. 
\end{itemize}

\subsubsection{Output layer}

In Section~\ref{sec:user-context} we explain how to compute a refined representation of users under different contextual situations. Based on this representation, we illustrate in Section~\ref{sec:knowledge-embedding} how to compute the refined representation of items. We now describe the output layer that calculates the predicted score (or probability).

The simplest way is simply using the inner product of vectors representing users and items, which is similar to MF \cite{koren2009matrix}. FM capture the pairwise interactions between embeddings \cite{rendle2010factorization}, therefore, using FM to compute the output is straight forward. Multi-layer perceptron (MLP) \cite{kruse2013multi} can capture the implicit interactions of features, therefore, it can be applied to compute the output. NFM \cite{he2017neural} is another strong baseline that combines FM and MLP, it is known for its ability to capture the second-order and non-linear interactions of features. We term these four variants as CA-KGCN-MF, CA-KGCN-FM and CA-KGCN-MLP, CA-KGCN-NFM respectively. The effects of them will be compared in Sections~\ref{sec:result_rating} and ~\ref{sec:result_ranking}.

\section{Experiment setup}
In this section, we evaluate the proposed CA-KGCN on three real-world datasets to verify CA-KGCN's ability in rating prediction and ranking prediction, which are two conventional methods for evaluating recommendation \cite{gunawardana2015evaluating}. Detailed statistics about the three datasets are presented in Table~\ref{tab:statistics}.

\begin{table}[]

\centering
\caption{Basic statistics of the three datasets. Note that non-item entities means the features of items.}
\label{tab:statistics}
\begin{tabular}{@{}lccc@{}}

\hline
Dataset                 & Frapp\'e & Yelp-CO & Yelp-WA \\ \hline
\#users                 & 957                     & 61469   & 49096   \\
\#items                 & 4082                    & 3198    & 3121    \\
\#interactions          & 96203                   & 142289  & 119576  \\
sparsity (users $\times$ items) & 97.54\%& 99.93\%& 99.92\%\\
\#Contextual factors    & 4                       & 3       & 3       \\
\#Contextual conditions & 98                      & 7       & 7       \\
\#Relation              & 5                       & 9       & 9       \\
\#Non-item entities     & 84                      & 223     & 223     \\
\#KG triplets           & 20410                   & 16079   & 16249   \\
Scale & 1-28752 (Number of interactions) & 1-5 & 1-5
\\ \hline
\end{tabular}
\end{table}

\subsection{Datasets}
The following datasets are utilized in our experiments. 

 \textbf{Frapp\'e:} This dataset is collected by Baltrunas et al. \cite{baltrunas2015frappe}. This dataset originated from Frapp\'e, a context-aware app recommender. There are 96303 logs of usage from 957 users under different contextual situations, 4082 apps are included in the dataset. Each contextual situation is composed of 8 contextual conditions: ``daytime'', ``weekday'', ``isweekend'', ``homework'',
       ``cost'', ``weather'', ``country'', ``city'' and each of them corresponds to a contextual factor. Since  ``weekday'' and ``isweekend'' contain almost the same information, we only keep ``isweekend'', which leads to slightly better results (in terms of AUC and F1).  ``cost'' is an attribute of app, we keep it as a relation in knowledge graph. In this dataset, logs in each country are concentrated in only a few cities, therefore, we only keep ``country''. As a result, we have 5 contextual factors: ``daytime'', ``isweekend'', ``homework (at home or at work)'', ``weather'', ``country''. The attributes of each app include: ``category'', ``downloads'', ``language``, ``price'', ``rating'', which indicates that there are 5 relations in knowledge graph.

 \textbf{Yelp\footnote[1]{https://www.kaggle.com/yelp-dataset/yelp-dataset}:} This dataset contains users' reviews on bars and restaurants in metropolitan areas in the USA and Canada. There are 8635403 observations in the whole dataset, due to the limited capacity of machine available\footnote[2]{Dell Latitude 7310, Intel® Core™ i5-10310U CPU @ 1.70GHz × 8 on Ubuntu 18.04.5 LTS}, we select the observations from Colorado and Washington. The number of observations from Colorado and Washington are close to that of the dataset Frapp\'e. 
 In the rest of this paper, they are noted as Yelp-CO and Yelp-WA respectively. As contextual factors, we have extracted ``day of week'', ``time of the day'', ``alone\_or\_companion''. The dataset also contains rich information about items' attributes, which can be represented in a knowledge graph. The attributes that we have extracted include: ``city'', ``stars (average ratings given by other users)'', ``review count ( the number of reviews received)'',
       ``ambience'', ``outdoorseating'', ``goodformeal (the speciality of the bar or restaurant)'', ``restaurantsGoodforgroups'',
       ``wifi''.

\subsection{Parameter learning and baselines}
As a matter of fact, CA-KGCN can be utilized for predicting the explicit scores that users will potentially give to items and for predicting the ranking of items, which depends on the \emph{output layer.} 

In score prediction scenarios, the mean squared loss \cite{gunawardana2015evaluating} is commonly used to optimize parameters of models:
\begin{equation}
    \mathcal{L} = \sum_{(u,i,cs,\mathcal{G})}{(\hat{r}_{(u,i,cs,\mathcal{G})} - r_{(u,i,cs,\mathcal{G})})}^2 + \lambda {\left \| \Theta \right\|}_2^2
\end{equation}

where $\hat{r}_{(u,i,cs,\mathcal{G})}$ denotes the predicted rating of user $u$ gives to item $i$ given knowledge graph $\mathcal{G}$ under contextual situation $cs$; $r_{(u,i,cs,\mathcal{G})}$ is the ground truth rating; $\Theta$ indicates the parameters of the model.

In ranking prediction scenario, point-wise log loss is commonly used to optimize the parameters of models \cite{lian2018xdeepfm,wang2019knowledge1}. The loss function to be optimized is:

\begin{equation}
    \mathcal{L} = -\sum_{\small(u,i,cs,\mathcal{G})\in y^+}log\mathcal{F}(\hat{y}_{u,i,cs,\mathcal{G}})-\sum_{\small(u,j,cs,\mathcal{G})\in y^-}log(1 - \mathcal{F}(\hat{y}_{u,j,cs,\mathcal{G}})) +\lambda {\left \| \Theta \right\|_2}^2
\end{equation}

where ${(u,i,cs,\mathcal{G}) \in y^+}$ indicates the observed interactions in the original dataset and ${(u,j,cs,\mathcal{G}) \in y^-}$ indicates the non-observed interactions under the same contextual situation $cs$, $\mathcal{F}$ represents the \emph{Sigmoid Function}, $\lambda$ is the regularization parameter to reduce over-fitting. $y^-$ is necessary because in some datasets, existing data records are all positive interactions. 

We compare our CA-KGCN with the following baselines:

\begin{itemize}
    \item[$\bullet$] \textbf{MF} \cite{koren2009matrix}: This is the classic collaborative filtering method that only explores user-item pairs, it simply computes the inner product of vector representing user and vector representing item to make predictions. Other information such as users' context and knowledge graph related to items are not considered.
    \item[$\bullet$] \textbf{FM} \cite{rendle2010factorization}: This is a strong baseline that captures the second-order interactions of all information related to interactions between users and items. The information here includes users' characteristics, users' context and item features.
    \item[$\bullet$] \textbf{NFM} \cite{he2017neural}: This baseline is a variant of FM and further applies MLP to capture the non-linear interactions between input vectors. In this sense, FM is a special case of NFM.
    \item [$\bullet$] \textbf{Deepfm} \cite{guo2017deepfm}: is another strong baselines that combines neural networks and factorization machine to explore the high-order and non-linear interactions.
    \item [$\bullet$] \textbf{LightGCN} \cite{he2020lightgcn}: is a collaborative filtering approach based on GCN. LightGCN models user-item interactions as a  bipartite graph. However, other information such as users' contexts and KGs related to items are not considered.
    
\end{itemize}

We note that the baselines and our proposed CA-KGCN are implemented through Pytorch \cite{paszke2019pytorch}. All the parameters are optimized by \emph{mini-batch Adam} \cite{kingma2014adam}. In order to decide the optimal hyper-parameters of models, we conduct a grid search: the learning rate is tuned on range $[5 \times 10^{-4}, 10^{-3}, 5 \times 10^{-3}, 10^{-2}, 5 \times 10^{-2}]$; the batch size is tuned on  $[128,256,512,1024]$; $L_2$ regularization term is tuned on range $[5 \times 10^{-4}, 10^{-3}, 5 \times 10^{-3}, 10^{-2}, 5 \times 10^{-2}, 10^{-1}]$; the dropout is tuned on range $[0,0.1,0.2,0.3,0.4,0.5]$; the embedding size are set to be 128 for all models.

 \section{Performances on rating prediction} \label{sec:result_rating}

The data records in the dataset \emph{Yelp} are scaled from 1-5. Therefore these ratings explicitly indicate users' preferences to items. We use Yelp-CO and Yelp-WA in the rating prediction scenario.

\subsection{Evaluation protocols}
 We randomly split Yelp-CO and Yelp-WA into three parts: $80\%$ for training set, $10\%$ for validation, $10\%$ for test. Training set is used for learning the parameters of CA-KGCN; validation set is used for tuning the hyper-parameters of CA-KGCN; the reported performances are compared on the test set. To evaluate the performances of models, we adopt the convention metrics: root mean
squared error (RMSE) and mean absolute error (MAE). Note that, the smaller they are the more accurate the predictions are.

\subsection{Results}

\begin{table}[]
\caption{Performance comparison of baselines and CA-KGCN on Yelp-CO and Yelp-WA. The bold indicates the best performances and the best performances of the baselines are underlined. * indicates statistically significant improvement given by paired \emph{t-test} with $p < 0.05$.}
\label{tab:result_prediction}
\begin{tabular}{@{}ccccc@{}}
\toprule
\multirow{2}{*}{Model} & \multicolumn{2}{c}{Yelp-CO} & \multicolumn{2}{c}{Yelp-WA} \\ \cmidrule(l){2-5} 
                       & RMSE          & MAE         & RMSE          & MAE         \\ \midrule
MF                     &   1.311            &   1.063          &  1.396             &  1.142           \\

LightGCN &1.263&0.997&1.254&1.014 \\
FM                     &  1.115             &    0.840         &  1.193             &    0.941         \\

Deepfm &1.069&\underline{0.809}&1.176&\underline{0.934} \\

NFM                    &  \underline{1.047}             &    0.812         &     \underline{1.137}         &     0.959        \\

\midrule
CA-KGCN-MF    &1.131    &    0.896           &  1.231           &        1.071                  \\
CA-KGCN-MLP   &   1.213  &    0.998           &  1.294           &         1.035                \\
CA-KGCN-FM  & 0.986      &    0.773           &  1.047           &     0.928                       \\
CA-KGCN-NFM   & \textbf{0.961*}     &     \textbf{0.736*}          & \textbf{0.992*}            &    \textbf{0.876*} \\
 \bottomrule
\end{tabular}
\vspace{-0,2cm}
\end{table}

Table~\ref{tab:result_prediction} presents the experiment results of rating prediction performances (aka. RMSE and MAE) on Yelp-CO and Yelp-WA. We note that the aggregator here is the sum aggregator, the impact of aggregator will be compared in Section~\ref{sec:ablation}. From this table we have the following observations: 

(1) MF constantly achieves the worst prediction accuracy. This is because MF only considers the user-item pairs, ignoring the vital information such as users' contexts and items features. Another reason is that Yelp-CO and Yelp-WA are quite sparse. The sparsity of Yelp-CO and Yelp-WA is $99.93\%$ and $99.92\%$ respectively, which makes the prediction less accurate. LightGCN explores the high-order interactions of users and items, therefore, it achieves better results than MF; 

(2) FM, Deepfm and NFM outperform MF, this shows that capturing the interactions between features helps to improve the accuracy of prediction. Besides, Deepfm and NFM outperform FM, this shows that leveraging users' context and item features also helps to deal with data sparsity issue. As for Deepfm and NFM, there is no single winner. Our interpretation is that the structure of the two models are similar and they both combine FM and MLP; 

(3) Our CA-KGCN approach constantly outperforms the baselines. We attribute the improvements of prediction accuracy to: (i) In the \emph{user-context embedding layer} of CA-KGCN, users and their contexts are modeled in a graph and users' preferences are modeled across their contexts. We model users' attention to different contextual factors, as a result, the vector representing users are adapted to their contexts; (ii) Besides, we leverage the knowledge graph related to items, the \emph{knowledge graph embedding layer} aggregates the information from item features to get a refined representation of items. In a word, the results show that leveraging users' contextual situations and the semantic information in the knowledge graph related to items do improve the accuracy of recommendations; 

(4) The last four rows present the performances of different variants of CA-KGCN. One interesting observation is that CA-KGCN-MLP achieves worse result than CA-KGCN-MF. This shows that CA-KGCN can get refined representation of users and items, as a result, simple inner product of user vector and item vector can make good predictions. Another potential reason is that we only adopt one layer of MLP, which limits its ability to model the non-linear interactions of input vectors; 

(5) From the last four rows, it can be observed that the choice of output layer influences the performance of CA-KGCN. Clearly, combining NFM and CA-KGCN achieves the best performances. In CA-KGCN-MF, the prediction score is computed simply by the inner product of the vector representing user and vector representing item (like in MF). However, CA-KGCN-NFM combing the advantages of FM and MLP: FM can capture the second-order interactions of features and MLP can capture the non-linear interactions of features. As a result, the CA-KGCN-NFM achieves better expressiveness and can better model users' preferences under different contextual situations.

\section{Performances on ranking prediction}\label{sec:result_ranking}
The interactions in the dataset \emph{Frapp\'e} indicate the frequency of use, which belong to implicit feedbacks of users. Therefore, this dataset is more adapted to ranking prediction. 

\begin{table}[t] 

\caption{Performance comparison of baselines and CA-KGCN on Frapp\'e on AUC, F1, HR@10, HR@20, NDCG@10 and NDCG@20. The bold indicates the best performances and the best performances of the baselines are underlined. * indicates statistically significant improvement given by paired \emph{t-test} with $p < 0.05$.}
\label{tab:ranking_auc}
\begin{tabular}{@{}cllllll@{}}
\toprule
            & AUC & F1 &HR@10&HR@20&NDCG@10&NDCG@20\\ \midrule
MF          &  0.811   & 0.557 &  0.259             & 0.317            &   0.122            &      0.142  \\
LightGCN &0.849&0.714&0.463&0.524&0.249&0.252\\
FM          & 0.855   & 0.743 &  0.537            &     0.648        &       0.339        & 0.371   \\
Deepfm &0.859&0.754& 0.596&0.722&0.365&0.404 \\
NFM         &\underline{0.881}     &\underline{0.760}  &  \underline{0.611}             & \underline{0.729}           &          \underline{0.386}     &   \underline{0.413}  \\ \midrule
CA-KGCN-MF  &  0.861   &    0.735& 0.541              & 0.675           &    0.329           &   0.365\\
CA-KGCN-FM  &    0.939 &    0.822&  0.633            &  0.744           &      0.417         & 0.426\\
CA-KGCN-MLP & 0.863   &  0.738 &   0.601            &     0.724        &     0.373          &   0.409 \\
CA-KGCN-NFM &   \textbf{0.942*}  &  \textbf{0.831*} &  \textbf{0.650*}             &   \textbf{0.772*}          & \textbf{0.453*}              &      \textbf{0.480*}    \\ \bottomrule
\end{tabular}
\vspace{-0,1cm}
\end{table}

\subsection{Evaluation protocols}
The leave-one-out evaluation protocol is a widely used to quantify the performances of RSs \cite{chen2019bayesian,he2017neural,he2017neural1,yuan2016lambdafm}, we randomly select one transaction as test set and the remaining as the training set. In order to learn a robust model, for each observed interaction we randomly sample 2 items that user has not interacted under contextual situation $cs$. Models are evaluated by the following way: (1) We use AUC and F1 to evaluate the click-trough-rate prediction, the larger AUC and F1 are, the better the models are; (2) We then evaluate models in the top-K recommendation scenario. We use the trained models to recommend K items for the positive interactions in the test set: Hit Ratio (HR) and Normalized Discounted Cumulative Gain (NDCG) are the conventional evaluation metrics in ranking prediction. We note that the ground truth in the testing stage is no longer the list of items associated with this user only, but the relevant items associated with users under the specific contextual situation. HR@K indicates whether the tested items are among the top-K recommended items and NDCG@K quantifies the position of the tested items \cite{shi2019deep}. The larger HR@K and NDCG@K are, the better the performance. In the experiment, we report the results of top-10 and top-20 recommendations on the test set. 

\subsection{Results}

Table~\ref{tab:ranking_auc} presents the performances comparison of CA-KGCN and baselines on AUC, F1 and top-K recommendations (HR@K and NDCG@K). Note that the aggregator here is the sum aggregator, the impact of aggregator will be compared in Section~\ref{sec:ablation}. We have the following observations: (1) Like in the rating prediction, MF achieves the worst results. This further confirms that only exploring the user-item pairs is not sufficient to model users' preferences. Besides, the high sparsity of the dataset Frapp\'e ($97.54\%$) is another reason why MF performs badly; (2) FM, Deepfm and NFM consistently outperform MF, since they can capture more useful information to model users' preferences; (3) CA-KGCN consistently performs better than the baselines. Among the variants, CA-KGCN-NFM achieves the best results. This shows that CA-KGCN-NFM can make highly accurate ranking predictions; (4) Like in rating prediction, CA-KGCN-MF outperforms CA-KGCN-MLP. This observation further verifies that CA-KGCN can get better representations of users and items. We attribute this to the fact that CA-KGCN leverages users' context and the rich semantic relations in the knowledge graph related to items.

 \begin{table}[t]
\caption{Impacts of aggregators. }
\label{tab:ablation_aggregator}
\begin{tabular}{@{}ccccccc@{}}
\toprule
\multirow{2}{*}{} & \multicolumn{2}{c}{Yelp-CO} & \multicolumn{2}{c}{Yelp-WA} & \multicolumn{2}{c}{Frapp\'e}\\ \cmidrule(l){2-7} 
                       & RMSE          & MAE         & RMSE          & MAE & AUC&F1         \\ \midrule
  AVG &0.972 &0.741 & 0.997& 0.880&0.935 & 0.821\\
  CAT &0.968 & 0.739&0.993 &\textbf{0.880} &\textbf{0.945} &\textbf{0.836} \\
          SUM &\textbf{0.961} &\textbf{0.736} &\textbf{0.992} &0.876 &0.942 &0.831
                    \\ \bottomrule
\end{tabular}
\end{table}

 \section{Analysis} \label{sec:ablation}
 
In this section, we study the following questions: the effects of aggregators (SUM aggregator and CAT aggregator, see Section \ref{sec:methodology}); the effects of users' contexts and knowledge graphs related to items; what CA-KGCN can learn from the dataset. More specifically, we answer the third question through a case study.

\textbf{Impact of aggregators:} In section~\ref{sec:methodology}, we propose two aggregators to aggregate information from neighboring nodes: SUM agregator and CAT agregator (see Section \ref{sec:methodology}). The AVG means the representation of neighborhood is computed by directly averaging the neighboring nodes, which means that there is no need to compute $\pi_u^{cf}$ and $\pi_u^r$, we add this variant to verify that computing users' attention to contextual factors and relations in the knowledge graph related to items do benefit prediction accuracy. Results in Table~\ref{tab:ablation_aggregator} show that the AVG aggregator performs worse than SUM aggregator and CAT aggregator. This shows that CA-KGCN can capture users' attention to different factors (their contextual situations and item features) that may influence their choices. This also leads to the idea that CA-KGCN helps to explain recommendation by using the attention level learned, which will be discussed in the next section. Comparing SUM aggregator and CAT aggregator, there is no single winner, for example in the dataset Yelp-CO, SUM aggregator outperforms CAT aggregator in rating prediction; in the dataset  Frapp\'e, CAT aggregator outperforms SUM aggregator in ranking prediction. This leads to the conclusion that the choice of aggregator depends on the dataset and the actual mission.

 \textbf{Impact of users' contexts and knowledge graph related to items:} To further verify the effects of leveraging users' context and knowledge graph related to items. We eliminate the \emph{user-context embedding layer} and the \emph{knowledge graph embedding layer} in CA-KGCN. It can be observed from Table~\ref{tab:ablation_context_knowledge} that, compared with MF that is context-free and knowledge-free, CA achieves better performances because it models users' preferences with finer granularity, this also confirms that users' context can influence their preferences; KGCN performs better with the benefit of the rich semantic information contained in the knowledge graph related to items; CA-KGCN benefits from users' context and knowledge graph related to items at the same time.

 \begin{table}[]
\caption{Impacts of users' context, knowledge graph related to items. Note that CA indicates that we only leverage users' context, KGCN means that only knowledge graph related to items is considered and CA-KGCN leverages users' context and knowledge graph at the same time.}
\label{tab:ablation_context_knowledge}
\begin{tabular}{@{}ccccccc@{}}
\toprule
\multirow{2}{*}{} & \multicolumn{2}{c}{Yelp-CO} & \multicolumn{2}{c}{Yelp-WA} & \multicolumn{2}{c}{Frapp\'e}\\ \cmidrule(l){2-7} 
                       & RMSE          & MAE         & RMSE          & MAE & AUC&F1         \\ \midrule
  MF & 1.311& 1.063&1.396 &1.142 &0.811 &0.557 \\
  CA &1.041 & 0.859& 1.115& 0.836&0.907 & 0.800\\
  KGCN &0.997 &0.757 &1.142 & 1.003&0.911 &0.802 \\
  CA-KGCN &\textbf{0.961} &\textbf{0.736} &\textbf{0.992} &\textbf{0.876} &\textbf{0.942} &\textbf{0.831}
                    \\ \bottomrule
\end{tabular}
\vspace{-0,2cm}
\end{table}

\textbf{Case study} In Sections~\ref{sec:result_rating}
and ~\ref{sec:result_ranking}, we show that the CA-KGCN outperforms baselines both in rating prediction and in ranking prediction. CA-KGCN can capture users' attention to different contextual factors and item features. Therefore, CA-KGCN can generate recommendations adapted to users' context. In this section, we show that CA-KGCN can generate context-aware explanations .

\begin{figure*}[t]
\centering  
\subfigure[\footnotesize{Cluster of users according to users' attention to contextual factors}]{
\label{Fig.sub.4}
\includegraphics[width=0.26\textwidth]{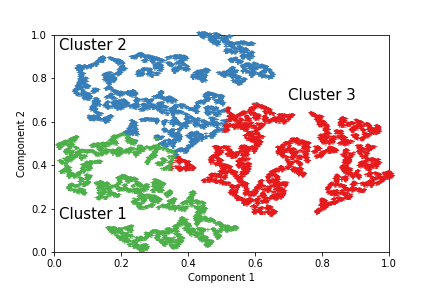}}
\subfigure[\footnotesize{Average of attention to contextual factors from cluster 1}]{
\label{Fig.sub.5}
\includegraphics[width=0.23\textwidth]{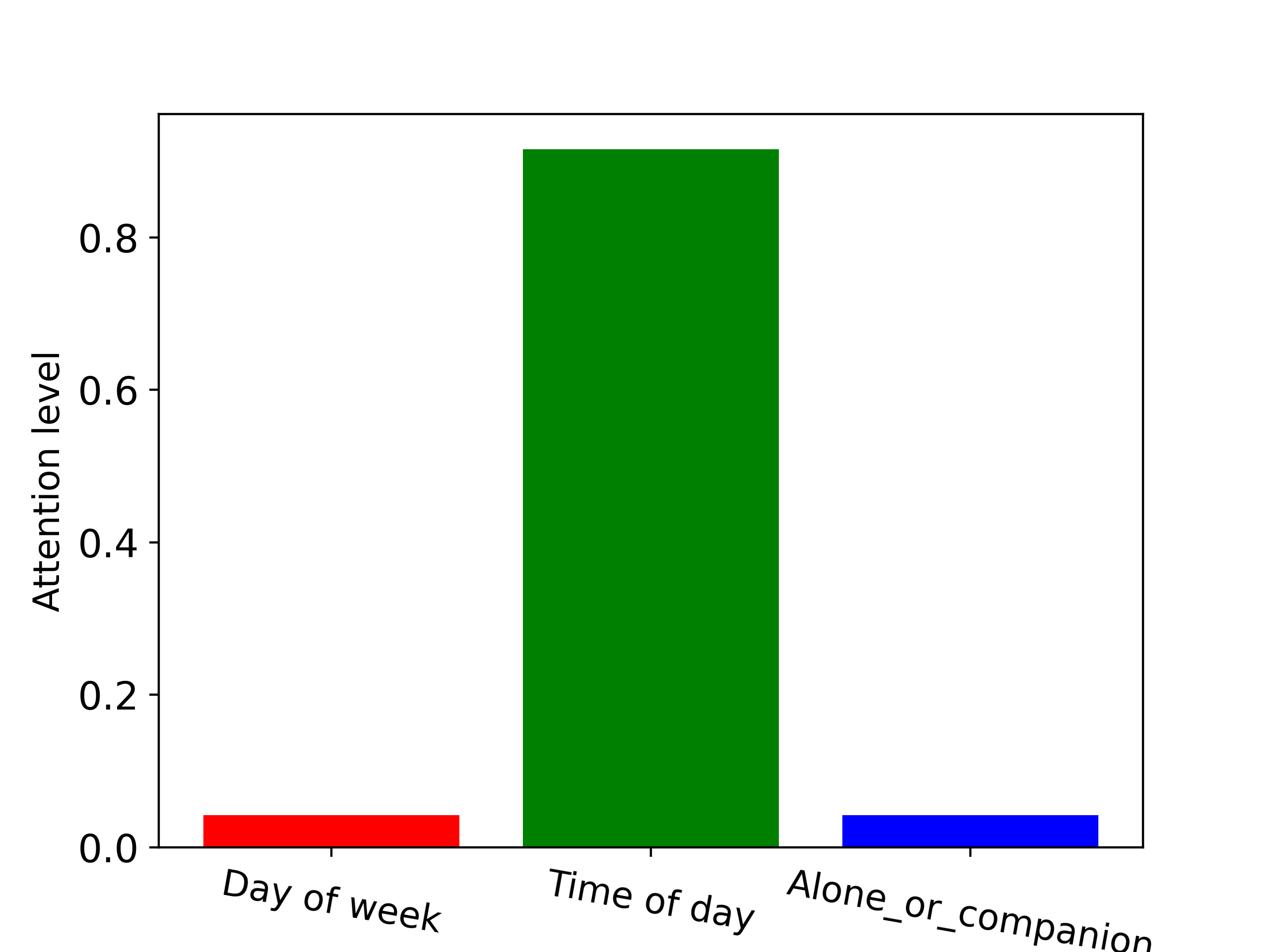}}
\subfigure[\footnotesize{Average of attention to contextual factors from cluster 2}]{
\label{Fig.sub.6}
\includegraphics[width=0.23\textwidth]{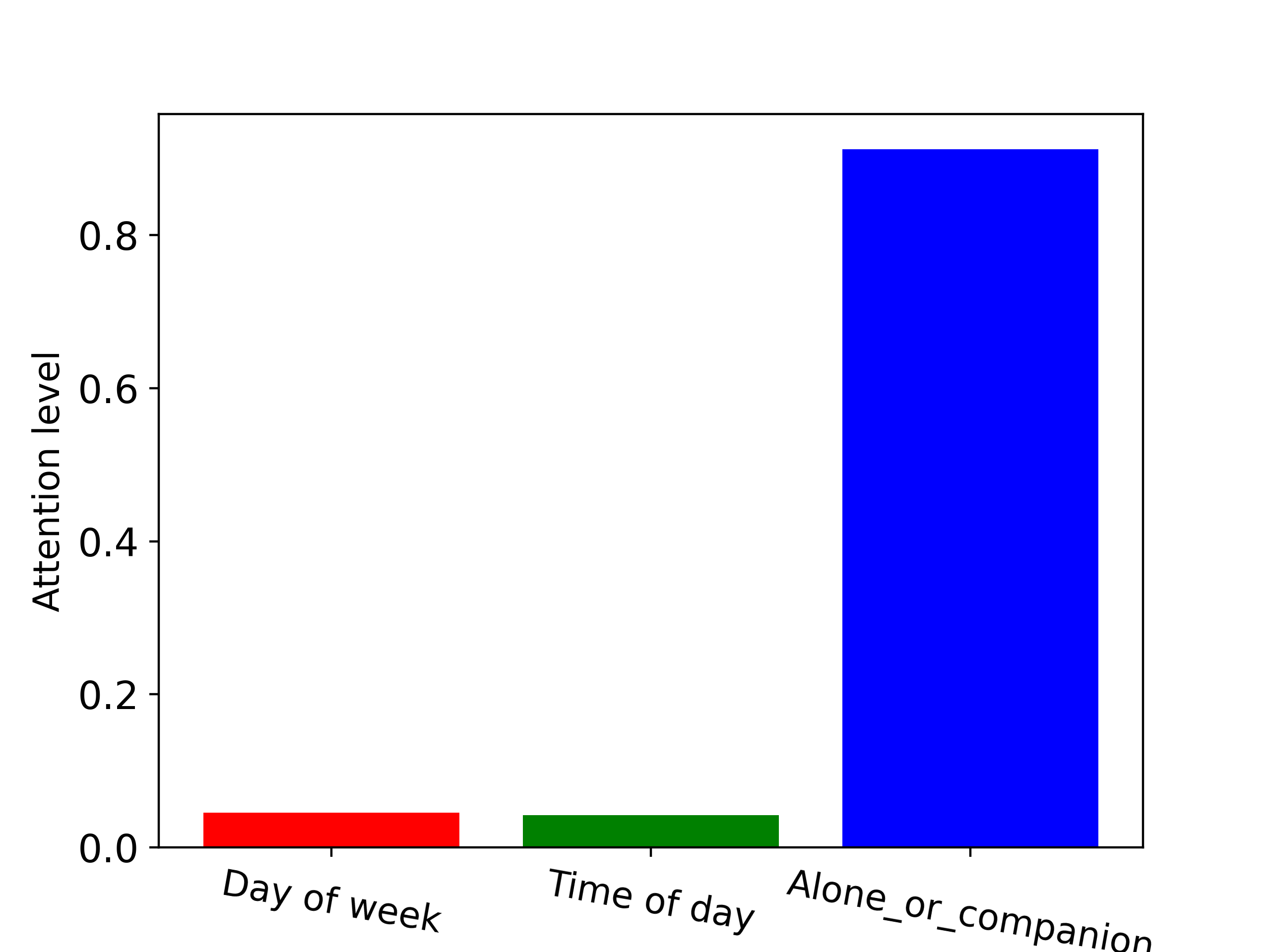}}
\subfigure[\footnotesize{Average of attention to contextual factors from cluster 3}]{
\label{Fig.sub.7}
\includegraphics[width=0.23\textwidth]{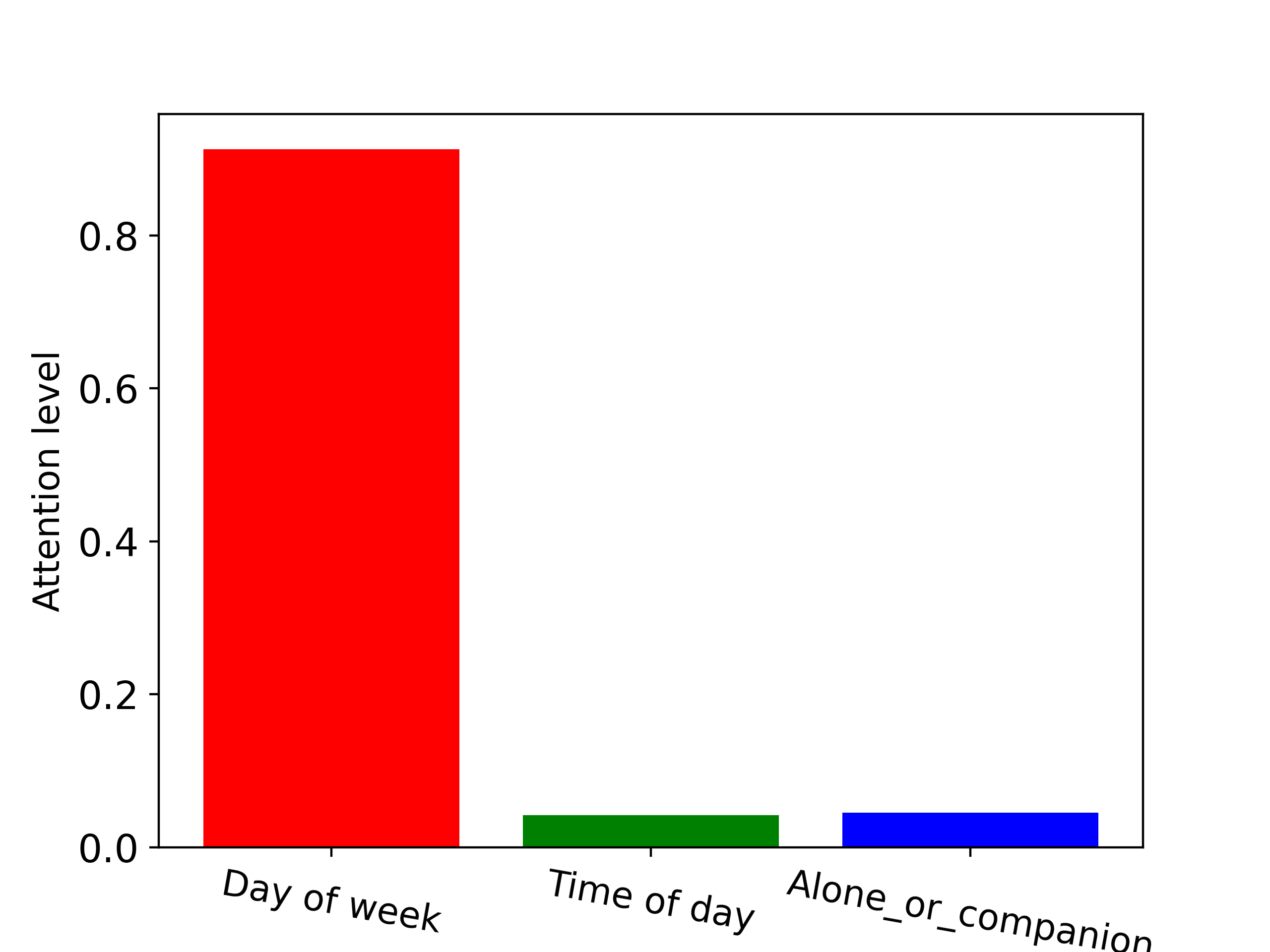}}
\subfigure[\footnotesize{User's attention to contextual factor}]{
\label{Fig.sub.1}
\includegraphics[width=0.29\textwidth]{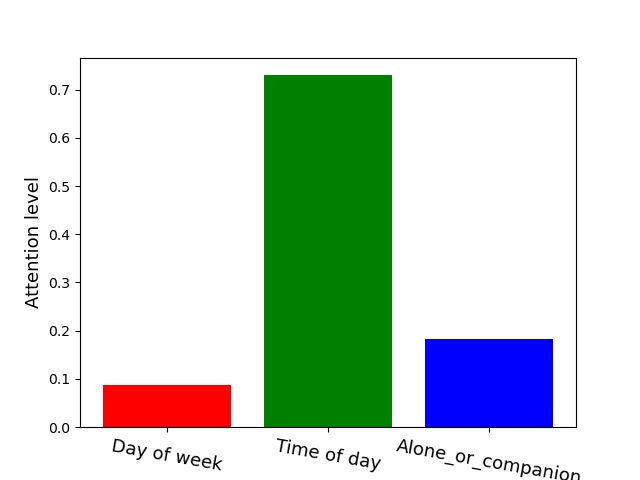}}
\subfigure[User's attention to different relations in the knowledge graph on a weekend night and with companion]{
\label{Fig.sub.2}
\includegraphics[width=0.31\textwidth]{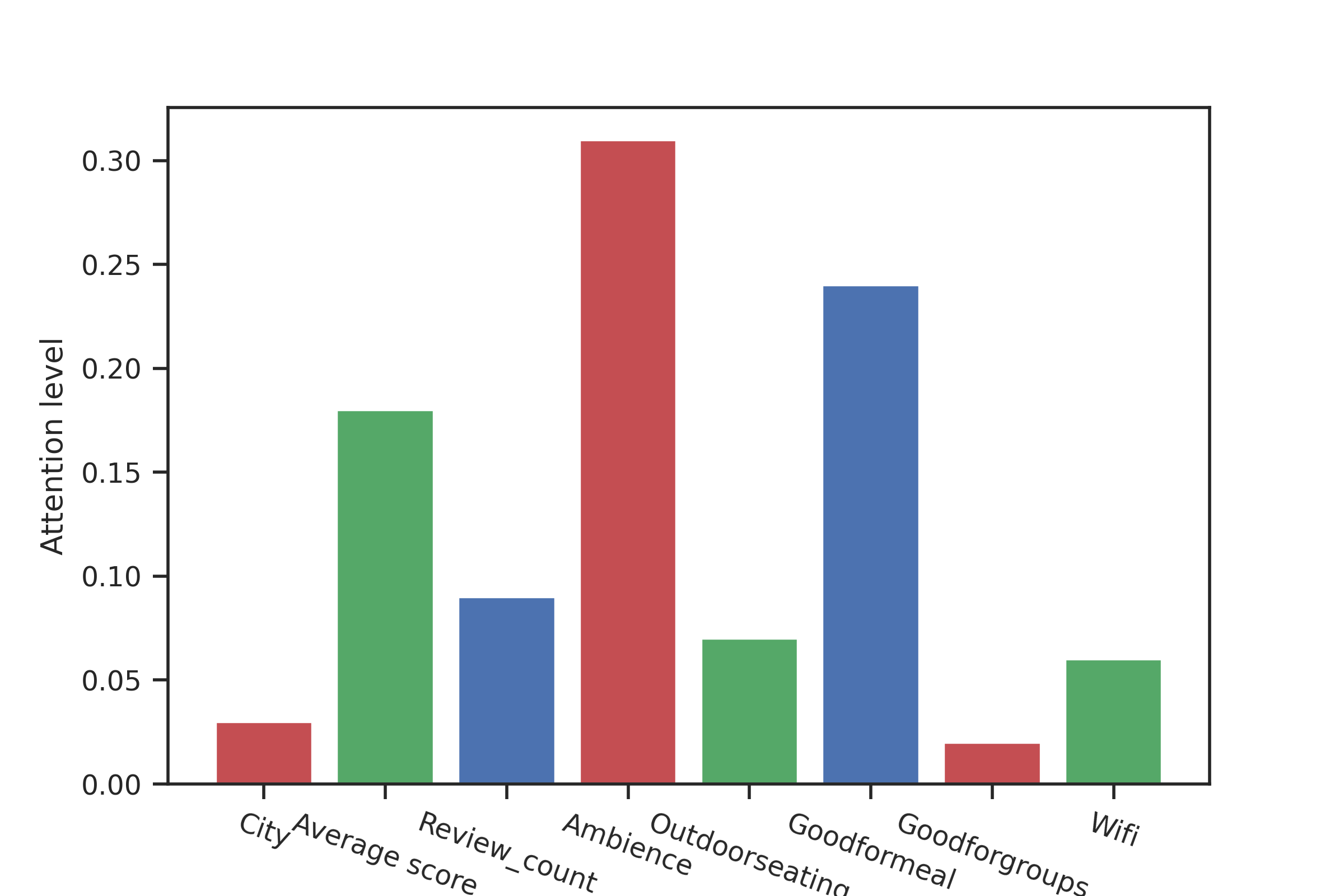}}
\subfigure[User's attention to different relations in the knowledge graph at a weekday noon and with companion]{
\label{Fig.sub.3}
\includegraphics[width=0.31\textwidth]{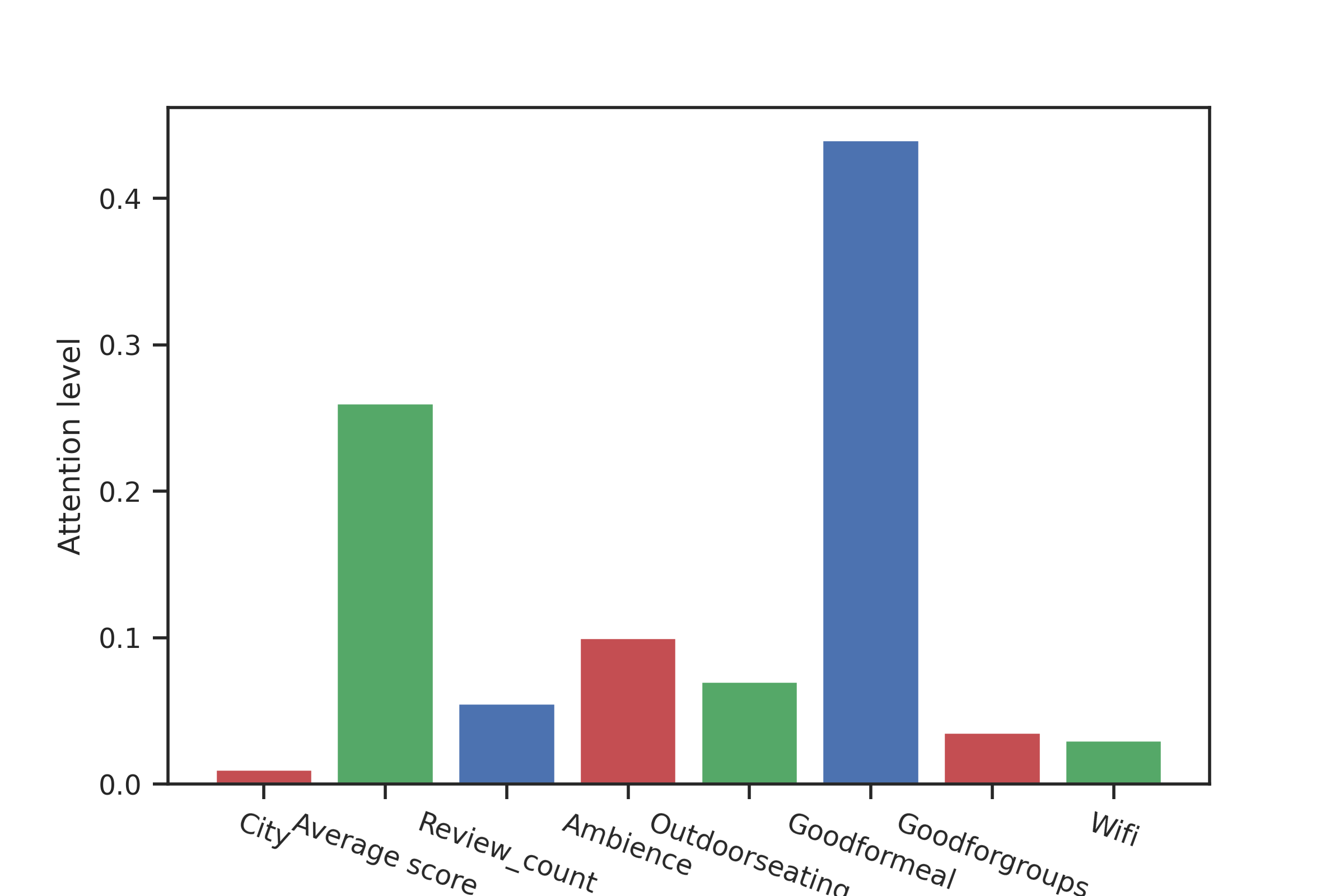}}
\vspace{-0,3cm}
\caption{Clusters of users and attention to contextual factors learned by CA-KGCN.}
\label{Fig.main}
\end{figure*}

As a case study, we select data records from the dataset Yelp-CO. Figure~\ref{Fig.main} shows the attention level and cluster of users learned by CA-KGCN. We first compute the users' (from test set) attention to different contextual factors. Then we use the results to represent each user. Three contextual factors are considered, therefore, each user is now presented by a vector of dimension 3. Next, we run K-means \cite{hamerly2003learning} to cluster the users. We have chosen K-means for its simpleness and effectiveness \cite{velmurugan2010computational}. Finally, we find that the number of clusters 3 suits the best. To visualize the clusters of users, we run t-SNE \cite{van2008visualizing} to visualize the clusters of users. t-SNE is chosen because t-SNE can preserve better the underlying information and structure of the data \cite{anowar2021conceptual}. A very first observation is that, users in the test set are clustered into three groups, as shown in Figure~\ref{Fig.sub.4}. Users from cluster 1 (see Figure~\ref{Fig.sub.5}) pay more attention to ``Day of week"; users from cluster 2 (see Figure~\ref{Fig.sub.6}) accord higher level of importance to ``Alone\_or\_companion"; users from cluster 3 (see Figure~\ref{Fig.sub.7}) pay more attention to ``Time of day". This confirms our assumption that users accord different level of importance to different contextual factors.

To verify how contexts influence users' preferences, we also zoom in on an user from cluster 1, Figure~\ref{Fig.sub.1} shows that this user thinks that the contextual factor ``Time of the day'' has slightly stronger effects on his choice of items. Another observation is that under different contextual situations, this user's attention to item features is different. Figure~\ref{Fig.sub.2} shows this user's attention level to different relations in the KG related to items on a weekend evening and with companion. Under this situation, ``Ambience'' and ``Goodformeal'' gain higher level of attention. Figure~\ref{Fig.sub.3} shows this user's attention level to different relations in the KG related to items at a weekday noon and with companion. This time, ``Goodformeal'' and ``Average score'' have larger weight. Figures~\ref{Fig.sub.2} and ~\ref{Fig.sub.3} illustrate that CA-KGCN can capture the effects of contextual factors and relations in knowledge graph. Therefore, CA-KGCN can model users' preferences with finer granularity, which leads to better performances.

We now propose an explanation scenario for this user. On a weekend evening, a recommended item $i$ specialized in providing dinner service and has a casual ambience  can be justified in the following way: ``It would be a good choice to have dinner in a restaurant with casual ambience on a weekend evening.'' At a weekday noon, a recommended item $i$ that is specialized in lunch service and has a average score of $4.1$ can be justified in the following way: ``It would be a good choice to have lunch in a restaurant whose average rating is $4.1$ at a weekday noon.'' Under the two contextual situations, recommendations are context-aware and explanations are also adapted to users' contextual situation.

In this section, we show that CA-KGCN benefits of leveraging users' contexts and knowledge graph related to items. By comparing the different aggregators, we show how each component of our framework benefits prediction. Through a case study, we show that CA-KGCN can learn users' attention to different contextual factors, which helps to cluster users and provide more personalized recommendations. We find that users can be clustered into three types according to their level of attention to contextual factors. We further zoom in on a single user and show that CA-KGCN can capture users' preferences under different contextual situations. Therefore, the explanations generated are also adapted to their situations.

\section{Conclusions and perspectives}

We now conclude and propose avenues for future work. In this paper, we propose \textbf{C}ontext-\textbf{A}ware \textbf{K}nowledge \textbf{G}raph \textbf{C}onvolutional \textbf{N}etwork (CA-KGCN), a framework that leverages users' contexts and knowledge graph related to items. This framework computes users' attention to their contexts and item features.  Experiments on three real-world datasets show that the variants of CA-KGCN can accurately predict ratings and rankings, indicating the ability of dealing with explicit feedback and implicit feedback respectively. We show through a case study that  CA-KGCN endows explicit meanings to embeddings, this helps to trace back how users' preferences are modeled. As a result, context-aware explanations adapted to users' context can be generated to justify recommendations. For future works, we propose the following avenues: (1) The dataset Yelp contains users' comments on the items they interacted before. Therefore, more contextual information can be extracted, we plan to explore users' comments to get a more precise description of users' contexts;
(2) We plan to carry out user studies to investigate users' preferences on explanations adapted to their contexts and explanations that are not context-aware; (3) We hope to carry out more study to investigate how aggregator influence results in different datasets.

\bibliographystyle{ACM-Reference-Format}
\bibliography{main}

\end{document}